\documentclass[aps,twocolumn]{revtex4}
\usepackage{eucal,amsmath,amssymb}
\usepackage[dvips,final]{graphicx}

\newcommand{\lp}{\left(}
\newcommand{\rp}{\right)}
\newcommand\DC{\Delta_{\mathrm{C}}}

\newcommand\Liou{{\mathcal L}}
\newcommand\Mfrak[1]{{{\mathfrak M}\lp#1\rp}}

\newcommand\Vquant[1]{{\mathcal V}\lp#1\rp}
\newcommand\osclength[1]{\xi_{#1}}

\newcommand\phot{{a^\dag a}}

\newcommand{\comm}[2]{\left[#1,#2\right]}
\newcommand{\ket}[1]{\left\vert #1\right\rangle}
\newcommand{\bra}[1]{\left\langle #1\right\vert}
\newcommand{\braket}[2]{\left\langle #1\left|\vphantom{{#1}{#2}}\right.#2\right\rangle}
\newcommand\avr[1]{\left\langle #1 \right\rangle}

\newcommand{\abs}[1]{\left\vert #1 \right\vert}

\newcommand{\omrec}{\omega_\text{rec}}

\newcommand\Ecal{{\mathcal E}}
\newcommand\Xcal{{X}}

\newcommand\Hmode{{-\DC\,\phot+i\eta\lp a^\dag-a\rp}}

\begin{document}

\title{Cavity nonlinear optics with few photons and ultracold quantum particles}

\author{Andr\'as Vukics}
\email{andras.vukics@uibk.ac.at}
\author{Wolfgang Niedenzu}
\author{Helmut Ritsch}
\affiliation{Institute for Theoretical Physics, University of Innsbruck, Technikerstr.~25, A-6020 Innsbruck, Austria}

\begin{abstract}
The light force on particles trapped in the field of a high-\(Q\)
cavity mode depends on the quantum state of field and particle.
Different photon numbers generate different optical potentials and
different motional states induce different field evolution.
Even for weak saturation and linear polarizability the induced
particle motion leads to nonlinear field dynamics.
We derive a corresponding effective field Hamiltonian containing all
the powers of the photon number operator, which predicts nonlinear
phase shifts and squeezing even at the few-photon level.
Wave-function simulations of the full particle-field dynamics confirm
this and show significant particle-field entanglement in addition.
\end{abstract}


\maketitle 

It is commonly assumed that offresonant interaction of coherent light
with a linear polarizable medium only creates linear phase shifts and
coherent states of the scattered light.
This is especially the case for weak light fields involving only few
photons, where generating any nonlinear phase shift is a serious
challenge \cite{shen67} while such devices would be highly desirable
in photonics or quantum information applications \cite{immamoglu97}.
One possibility to generate nonlinear phase shifts even at the single
photon level was demonstrated via resonant strong coupling of an
atomic transition and a photon in a high-\textit{Q} cavity
\cite{turchette95}.
Recently it was noted that nonclassical light fields can also emerge
from linear scattering off a weakly excited medium, if its motional
state has genuine quantum properties \cite{mekhov07a}.
Here we combine those two ideas and study the effective nonlinear
cavity field dynamics generated from the photon number dependence of
light forces on the quantum particle motion in the cavity-enhanced
optical potential \cite{gangl00a}.

In a prototype setup an ultracold particle is placed in the optical
potential generated by a cavity field mode.
For large detuning between atomic and cavity resonance spontaneous
emission is small and the cavity field simply generates a
photon-number dependent trapping potential.
For any superposition state involving several photon numbers, as
e.g.~a coherent state, the localization of the particle thus differs
for each photon number.
As the particle position spread in turn determines effective
refractive index and field phase shift this implies effective
nonlinear field evolution.
Experimentally a closely related nonlinear effect was found in a
pioneering setup in Berkeley \cite{gupta07}.
As several similar setups coupling a BEC to a high-\textit{Q} cavity
mode have just been realized \cite{brennecke07}, further
detailed experimental studies will emerge soon.

Naturally the coupled particle-field dynamics generates entanglement,
which partly persists even in a steady state, where each photon number
is correlated with a different particle state.
Interestingly, even after tracing over the particle degrees of freedom
the field exhibits nonclassical properties and its dynamics can be
approximated by a simple effective field Hamiltonian.
We study its central predictions, and check its validity by a full
particle-field wave-function simulation.

\begin{figure}
\centering
\includegraphics[width=.9\columnwidth]{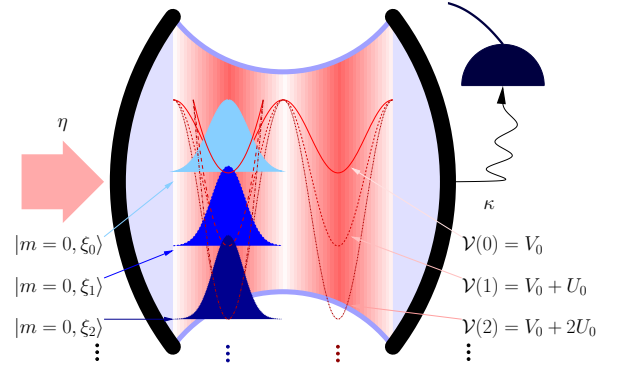}
\caption{
(Color online)
Scheme of the system with a quantum particle in the field of a driven
cavity.
A single driven standing wave mode generates optical potentials of
different depth depending on the photon number.
We also display the corresponding particle ground-state wavefunctions.
}
\label{fig:scheme}
\end{figure}

Our prototype CQED system is depicted in Fig.~\ref{fig:scheme}.
Following standard assumptions for a particle moving in 1D along the
axis of a far red detuned single cavity mode, we start from a
Jaynes-Cummings Hamiltonian with eliminated internal state(s)
\cite{domokos03,vukics05b} to get (\(\hbar=1\)):
\begin{equation}
\label{eq:Ham}
H=\frac{p^2}{2\mu}+(V_0+U_0{a^\dag a})f(Kx) \Hmode.
\end{equation}
Here \(x\), \(p\), and \(\mu\) are the particle position and momentum
operators and its mass; \(a\) is the field operator, \(K\) is the mode
wave number, \(\DC=\omega-\omega_{\mathrm{C}}\) is the cavity detuning
(\(\omega\) is the laser, \(\omega_{\mathrm{C}}\) is the mode
frequency), \(\eta\) is cavity pump amplitude.
\(V_0\) describes an additional c-number potential along the cavity
axis, it ensures a particle bound state in the absence of cavity
photons, but can be neglected for larger photon numbers.
A central parameter here is \(U_0\), the optical potential depth per
photon, which also gives the cavity frequency shift per particle and
is proportional to the particle susceptibility \cite{vukics05b}.
As we consider small cavities, photon loss is important even for good
mirrors and will be described by the standard Liouvillean
\cite{gardiner} \(\Liou\rho=\kappa\lp2a\rho
a^\dag-\comm{a^{\dag}a}{\rho}_+\rp\), where \(\kappa\) is the mode
linewidth. In the following \(\Mfrak{H}\) stands for a \emph{Master
  equation} defined by Hamiltonian \(H\) and this Liouvillean.
%

The most interesting part of the dynamics arises from the second term
of the Hamiltonian (\ref{eq:Ham}) describing dispersive particle-field
interaction.
Its effect can clearly be seen from an expansion of the wavefunction
in a product basis \(\ket\Psi=\sum_n\Psi_n\ket n\ket{\phi_n}\), where
\(\ket n\) is the \(n\)-photon field state and \(\ket{\phi_n}\) a
particle wavefunction, which will evolve subject to the potential
\(\Vquant{n}\,f(Kx)\)
\footnote{We define the potential strength \(\Vquant{\Xcal}\equiv
  V_0+U_0\,\Xcal\), and from this the oscillator frequency
  \(\Omega(\Xcal)^2\equiv4\omrec\Vquant{\Xcal}\) and the
  characteristic length
  \((K\osclength\Xcal)^4\equiv\omrec/\Vquant{\Xcal}\). Here
  \(\omrec\equiv\hbar K^2/(2\mu)\) is the recoil frequency, which
  gives the elementary timescale of the atomic motion on a length-scale
  of \(K^{-1}\). \(\Xcal\) may stand for an operator or a c-number:
  hence e.g.~\(\Vquant\phot\) represents a \emph{quantum} potential
  strength, while \(\Vquant n\) is the potential created by the
  n-photon Fock state of the mode. We also use the notation
  \(\ket{m,\osclength{}}\) for the \(m\)th harmonic oscillator
  eigenstate corresponding to oscillator length \(\osclength{}\)}.
Naturally, this creates entanglement between the particle and field on
the timescale of the particle motion, characterized by \(\omrec\).
The pump term proportional to \(\eta\) in the Hamiltonian
(\ref{eq:Ham}) and the photon loss described by quantum jumps
(application of operator \(a\) on the state vector) on the other hand
mix the different evolution branches and thus tend to reduce the
entanglement.

The fullness of this very intricate stochastic dynamics can be
captured only by a simulation performed on the product basis of the
mode-particle Hilbert space, which leads to high dimensionality even
for a single particle \cite{vukics05a,vukics07a}.
Fortunately to understand the central physics, such a brute force
approach is not necessary and we will present a systematic method for
identifying a much lower dimensional but sufficient subspace in this
immense Hilbert space.
It is spanned by separable particle-field states, which represent the
full quantum trajectories very well after an initial transitional
period.
The idea should be generally applicable in situations where two
quantum mechanical systems interact and one of them is dissipative ---
a situation ubiquitous in cavity QED, but also e.g.~in a multi-level
atom where the dissipative internal degree of freedom and the motion
are coupled by a spatially dependent pump.

As the cavity field is well represented by a standing wave cosine mode
with wave number \(K\), \(f(Kx)=\cos^2(Kx)\), the Hamiltonian commutes
with the parity operator so that the dynamics does not mix the
symmetric and antisymmetric subspaces of the particle Hilbert space
and we will restrict ourselves to the symmetric part only.
To further simplify the mathematics in the following we add a
sufficiently strong classical part to the cosine potential, so that
with \(V_0,U_0<0\) we are allowed a harmonic approximation
\(f(Kx)\approx1-(Kx)^2\).
At the same time we redefine the detuning \(\DC-U_0\to\DC\) and
\(\Vquant{\Xcal}\equiv\abs{V_0}+\abs{U_0}\,\Xcal\).

Let us now construct a subspace spanned by separable states, which
works well in the regime of moderate coupling in the sense that the
expectation value of the projector to the subspace is close to unity
for most of the time on a trajectory.
As a particle with a \emph{fixed} state vector cannot get entangled
with the field, we get a coherent steady state in the mode.
%
%
%
%
Therefore, if we fix the particle in a harmonic oscillator eigenstate,
the corresponding stationary field is a coherent state \(\ket\alpha\),
which can be simply calculated as the particle merely causes a
frequency shift of the mode by \(\abs{U_0}K^2\avr{x^2}\).
This coherent state in the mode creates an additional average
potential for the particle reading \(\Vquant{\abs\alpha^2}\,(Kx)^2\).
For the combined potential we then can again calculate the
corresponding particle eigenstate.
Iterating this process (or simply solving a corresponding nonlinear
equation for \(\alpha\)) we find the subspace
\begin{equation*}
\Ecal^\text{(coh)}\equiv\text{span}\left\{\ket{\alpha_m}\ket{m,\osclength{\abs{\alpha_m}^2}}\right\}_{m\in\mathbb{N}},
\end{equation*}
where the particle state is a harmonic oscillator eigenstate with
oscillator length \(\osclength{\abs{\alpha_m}^2}\) determined together with
\(\alpha_m\) self consistently.

\begin{figure}
\centering
\includegraphics[width=.7\columnwidth]{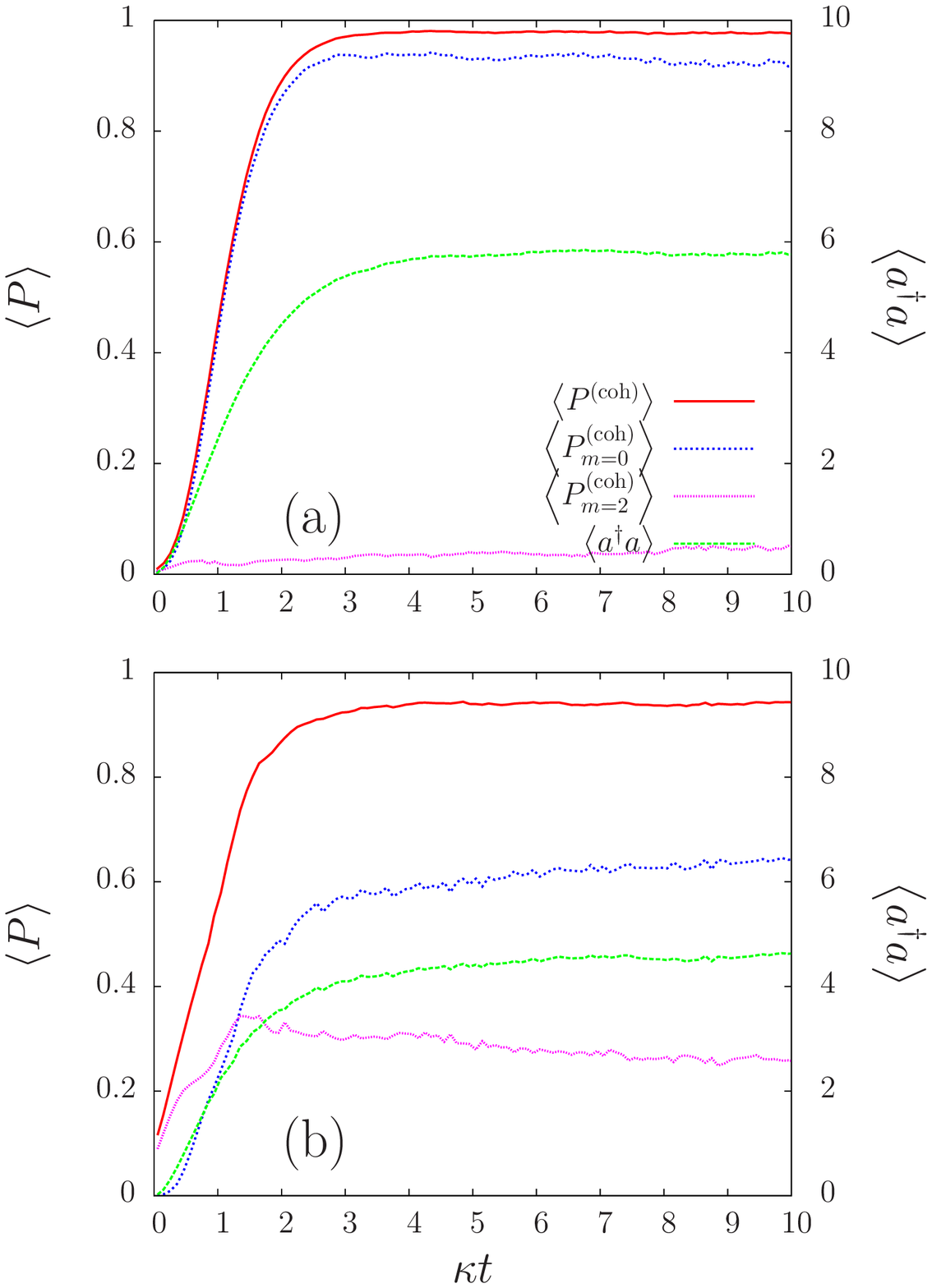}
\caption{
(Color online) 
System time evolution using an ensemble of 300 trajectories started
from a product state of vacuum for the mode and (a)
\(\ket{0,\osclength{0}}\), (b)
\(\lp\ket{0,\osclength{0}}+\ket{2,\osclength{0}}\rp/\sqrt2\).
After a transitional period during which the photon number grows from
0 to its quasi steady-state value the system arrives into the subspace
\(\Ecal^\text{(coh)}\), where it stays from then on, so that
\(\avr{P^\text{(coh)}}\) is close to one.
Parameters: \((\kappa,V_0)=(10,-100)\omrec\),
\((\DC,\eta,U_0)=(0,2.5,-10)\kappa\).
This \(U_0\) value represents a moderate coupling regime for our
purposes.
\(P^\text{(coh)}_{0,2}\) represent the projectors to only the states
corresponding to \(m=0,2\).
As we see, the population in these two states in the quasi
steady-state differs from that in the initial condition (\(50-50\%\)):
the \(m=0\) state has higher population, in which the cavity cooling
effect \cite{horak97} is manifested at work.
}
\label{fig:overlapsAlpha}
\end{figure}

In Fig.~\ref{fig:overlapsAlpha} we demonstrate that
\(\Ecal^\text{(coh)}\) is a very suitable subspace to represent the
full coupled particle-field dynamics in the moderate coupling regime.
This is done by calculating the expectation value of the subspace's
projector \(P\) for a state obtained from Monte Carlo wave-function
simulations of the full coupled dynamics with a dimension of
\(\approx2000\) \cite{vukics07a}.

Being based on self-consistent product states, the above method for
obtaining \(\Ecal^\text{(coh)}\) contains essential parts of the
nonlinear field dynamics but completely disregards the expected
particle-field entanglement.
Hence this ansatz (and the subspace \(\Ecal^\text{(coh)}\)) is
unsuitable for stronger coupling where we have to use an extended
approach involving a somewhat larger set of separable states.
While the particle states will still be harmonic-oscillator
eigenstates in self-consistent average potentials, the corresponding
mode states have to be determined by a new Master equation based on an
effective nonlinear Hamiltonian invoking all the powers of the photon
number operator.
Note that although it is eliminated from the dynamics it is the very
\emph{quantum nature} of the motion of the \emph{linearly polarizable}
particle, which creates this nonlinearity.
Besides giving a very good intuition to the underlying dynamics, it is
a significant boon of this approach that we are left with a dynamics
to be solved solely on the mode Hilbert space, with a much smaller
dimension.
We start the derivation by casting the Hamiltonian (\ref{eq:Ham}) into
the form
\begin{equation}
\label{eq:HamHarm}
H_\text{HO}=\Omega\lp\phot\rp\lp b^\dag b+\frac12\rp \Hmode.
\end{equation}
The first two terms of that Hamiltonian in the harmonic-oscillator
approximation can be conveniently expressed using the ladder operator
defined as \(b\equiv(\xi_\phot^{-1}\otimes x+i\xi_\phot\otimes
p)/\sqrt2\), so that it is an operator on the complete mode-particle
Hilbert space.
Note for future reference that \(b,\,b^{\dagger}\) still obey the
usual commutation relations.
Recall that the frequency \(\Omega(\phot)\) is an operator on the mode
Hilbert space.

Consider now the following basis:
\(\ket{n,m}\equiv\ket{n}\ket{m,\osclength{n}}\), where \(\ket n\) is
the \(n\)th Fock state of the mode, while \(\ket{m,\osclength{n}}\) is
again the \(m\)th harmonic oscillator eigenstate corresponding to the
oscillator length determined by the state of the mode (c.f.~also
Fig.~\ref{fig:scheme}).
Due to this dependence the basis is \emph{not} a direct product of
bases in the subsystems' Hilbert spaces.
Nevertheless, it is an orthogonal basis, since the Fock states for
different \(n\) are orthogonal, while for a given \(n\) the
\(\ket{m,\osclength{n}}\) states are orthogonal for different \(m\)
since they have the same oscillator length.
We now partition the Hilbert space to subspaces with a given \(m\)
\(\Ecal_m\equiv\left\{\ket{n,m}\right\}_{n\in\mathbb N}\) with
projectors \(P_m\equiv\sum_n\ket{n,m}\bra{n,m}\).

Let us project the Master equation \(\Mfrak{H_\text{HO}}\) to the
subspace \(\Ecal_m\).
It is an invariant subspace of the first (\(H^\text{(int)}\)) and
second terms of the Hamiltonian (in fact, the \(\ket{n,m}\) states are
eigenstates), so that e.g.~\(H^\text{(int)}=\sum_mP_mH^\text{(int)}P_m\).
This is not true, however, for the pump term (third term of the
Hamiltonian (\ref{eq:HamHarm})) and the Liouvillean, since operator \(a\) mixes
the subspaces with different \(m\).
Indeed, \(a\ket{n,m}\propto\ket{n-1}\ket{m,\osclength{n}}\), and
\(\braket{m',\osclength{n-1}}{m,\osclength{n}}\neq0\) for all \(m'\),
therefore \(\bra{n-1,m'}a\ket{n,m}\neq0\).
In a first approximation we neglect the coupling between the
\(\Ecal_m\) subspaces by the operator \(a\) and assume
\(H\approx\sum_mP_mHP_m\) for the complete Hamiltonian
(\ref{eq:HamHarm}) and the Liouvillean.
A mathematical justification of this is that \(\osclength{n}\) is only
slowly varying with \(n\), so that the corresponding eigenfunctions
are almost orthogonal.
All approximations are \emph{a posteriori} confirmed by numerical
tests.
Within this approximation the dynamics can be solved separately in the
different \(\Ecal_m\) subspaces.
Note that the projected field operator is approximated
\(P_maP_m=\approx\sum_m\sqrt{n+1}\ket{n,m}\bra{n+1,m}\), and that this
approximation is not crucial, but it yields that the usual \(a\)
operator will appear in the projected Master equation
\(\Mfrak{H_m}\).
Within this framework we can find the steady-state density operator in
the subspace \(\Ecal_m\), which has the dimension of only the field
Hilbert space, of the form
\(\rho_m^{\text{(ss)}}\equiv\sum_{n,n'}\rho_{m;n,n'}^{\text{(ss)}}\ket{n,m}\bra{n',m}\),
where \(\rho_{m;n,n'}^{\text{(ss)}}\) is \emph{formally} a density
matrix only on the mode Hilbert space. It is the steady-state solution
of \(\Mfrak{H_m}\) with effective Hamiltonian
\begin{multline}
\label{eq:HamSqrt}
H_m\equiv\sqrt{\omrec\lp\abs{V_0}+\abs{U_0}\phot\rp}\lp2m+1\rp\\\Hmode.
\end{multline}
Note that for \(\omrec=0\) we recover the standard single mode
Hamiltonian linear in \(\phot\), which yields pure coherent states in
steady state.

\begin{figure}
\centering
\includegraphics[angle=-90,width=.9\columnwidth]{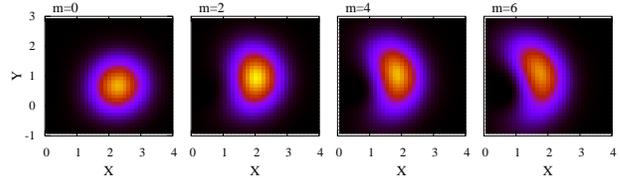}
\caption{
(Color online)
Wigner functions calculated from the stationary field density matrix
\(\rho_{m;n,n'}^{\text{(ss)}}\) for different vibrational particle
states \(m\).
Same parameters as for Fig.~\ref{fig:overlapsAlpha}, but \(\DC\) is
always adjusted to resonance, i.e. in the Hamiltonian
(\ref{eq:HamSqrt}) the term linear in \( \phot\) vanish.
While for \(m=0\) the state is almost coherent the Wigner function
becomes more and more banana-shaped with increasing \(m\).
}
\label{fig:WF}
\end{figure}

From a Hamiltonian (\ref{eq:Ham}) linear in \(\phot\), but containing
the particle-mode interaction, we have hence arrived at the
conditional Hamiltonian (\ref{eq:HamSqrt}) nonlinear in \(\phot\)
operating only on the field Hilbert space.
In Fig.~\ref{fig:WF} we illustrate the nonclassical states this
Hamiltonian generates. As striking example one obtains banana-shaped
Wigner functions for higher \(m\) indices.

To find a subspace which represents \(\rho_m^{\text{(ss)}}\) well,
we adopt an idea from the density matrix renormalization group method
\cite{schollwockreview}.
We diagonalize the density operator and take a (usually small) number
of eigenvectors corresponding to the leading eigenvalues --- in fact,
we introduce a cutoff at a given value \(\epsilon\ll1\) in the spectrum.
Let us remark here that for parameters we studied in practice,
\(\rho_m^{\text{(ss)}}\) was always very close to a pure state
(largest eigenvalue very close to unity).
We obtain the states
\(\ket{\Psi_{m,i}}\equiv\sum_n\Psi_{m,i;n}\ket{n,m}\), where \(i\)
indexes the different eigenvectors.
It runs between 1 and \(N_m\), which latter is a number depending on
\(\epsilon\) and it may be different for different \(m\) indices.

Our most salient approximation is that at this point we approximate
these states by separable states:
\begin{multline}
\ket{\Psi_{m,i}}\approx\ket{\phi_{m,i}}\ket{m,\osclength{\bra{\phi_{m,i}}\phot\ket{\phi_{m,i}}}}\equiv\ket{e_{m,i}},\\
\text{with}\;\ket{\phi_{m,i}}\equiv\sum_n\Psi_{m,i;n}\ket{n}.
\end{multline}
Note that the particle state has been taken to be an \(m\)th harmonic
oscillator eigenstate corresponding to the oscillator length
calculated from the average photon number of the mode state
\(\ket{\phi_{m,i}}\).
This approximation is on one hand again justified by the
aforementioned mathematical argument, but also by a physical argument:
during the dynamics the pump term and the quantum jumps mix the levels
of different \(n\), and a particle with finite mass cannot follow this
mixing immediately.
In addition it is \emph{a posteriori} justified by simulations.
For the sake of further intriguing the reader we note that the states
\(\ket{e_{m,i}}\) are not even in the subspace \(\Ecal_m\) to which we
projected our Master equation in the first place.

\begin{figure}
\centering
\includegraphics[width=.7\columnwidth]{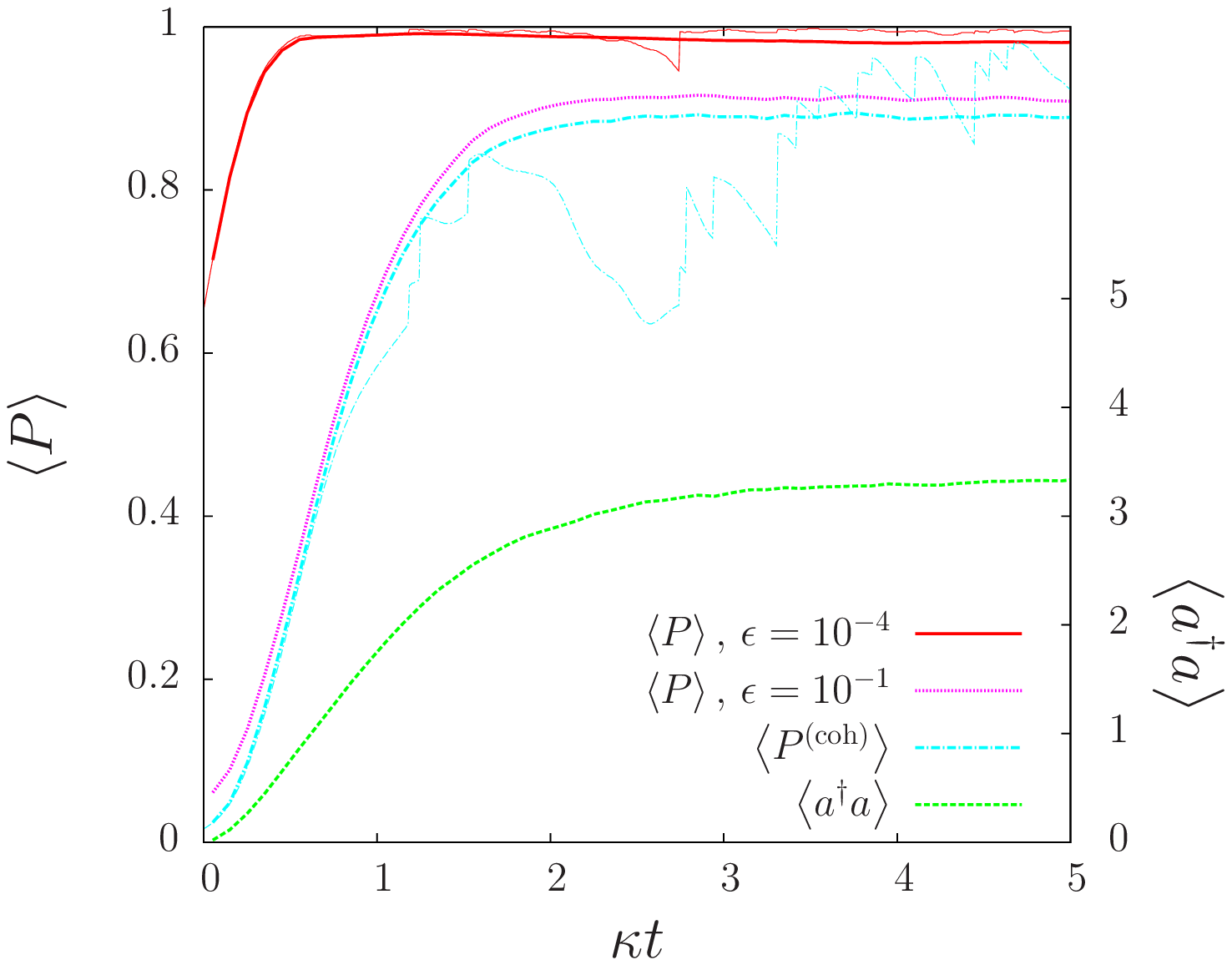}
\caption{
(Color online)
Projector expectation values in the strong coupling regime
\((\DC,\eta,U_0)=(0,2.5,-100)\kappa\).
Ensemble of 300 trajectories.
With \(\epsilon=10^{-1}\) \(N_m=1\) for all \(m\), while with
\(\epsilon=10^{-4}\) \(N_0=3,\,N_2=2,\,N_{m>2}=1\).
In both cases \(\avr{P}>\avr{P^\text{(coh)}}\), but indeed in the
second case \(\avr{P}\approx1\) after a transitional period.
In thin lines results from a typical \emph{individual} trajectory are
displayed to show that the quality of \(\Ecal^\text{(coh)}\)
fluctuates wildly on a single trajectory, while that of \(\Ecal\)
remains close to unity even in this case.
}
\label{fig:overlaps}
\end{figure}

We readily arrive at the subspace
\begin{equation*}
\Ecal\equiv\text{span}\left\{\ket{e_{m,i}}\right\}_{m\in\mathbb N, i=1\dots N_m}
\end{equation*}
spanned by separable states, for which the mode states are leading
eigenvectors of a density matrix, which is a steady-state solution of
\(\Mfrak{H_m}\) containing the square root of \(\Vquant\phot\) and
hence all the powers of the photon number operator.
In Fig.~\ref{fig:overlaps} we demonstrate that this is a very high
quality subspace even in the (for our purposes strong coupling) regime
where the formerly derived subspace \(\Ecal^\text{(coh)}\) breaks
down.
As expected, the quality of \(\Ecal\) can be increased by decreasing
the cutoff parameter \(\epsilon\).

\begin{figure}
\centering
\includegraphics[width=.8\columnwidth]{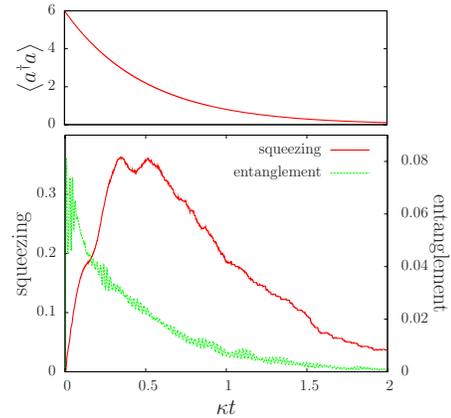}
\caption{
(Color online) 
Time evolution (300 trajectories) from an initial product state.
Parameters: \((\kappa,V_0)=(0.1,-60)\omrec\),
\((\DC,\eta,U_0)=(0,0,-100)\kappa\) --- the particle is chosen to be
very light in comparison to previous cases, so that it can be
influenced more easily by the field.
The arising field squeezing is measured by \(-\log\lp\lambda_s\rp\),
where \(\lambda_s\) is the smaller eigenvalue of the field
quadratures' correlation matrix.
The particle-field entanglement is measured by the negativity of the
density operator's partial transpose \cite{vidal02,vukics05a}.
}
\label{fig:empty}
\end{figure}

In Fig.~\ref{fig:empty} we demonstrate one more experimentally very
easily accessible situation, which clearly exhibits the
nonclassicality of the field generated by particle-field entanglement.
Here to even more distill the central physical effect, the cavity is
chosen to be unpumped, but a few-photon coherent state is prepared in
the mode, which then slowly leaks out.
The particle is started from the ground state wave packet of the
complete potential, so that the system is initially in a product
state.
Note that via the generated entanglement the particle is capable to
transform the initial coherent field state into a squeezed state.

We demonstrated that few photon nonlinear optics can be implemented
even by help of a linearly polarizable medium, if one includes spatial
dynamics of the medium.
In the low temperature limit where the medium has genuine quantum
properties, a coherent state input field then can get entangled with
the motional states.
Surprisingly this complex coupled dynamics can be described by an
effective nonlinear field Hamiltonian which can be tailored to
experimental needs.
As it operates only on a restricted Hilbert space it can be
efficiently simulated.
In our example we could identify an O(10) dimensional subspace of the
complete O(1000) dimensional Hilbert space to which the system
dynamics was confined with high probability.
While we have concentrated on a single quantum particle here, an
analogous dynamics will emerge in the many particle case if one
invokes only the lowest few collective excitations of the
ensemble. 
This case will also show a strongly increased nonlinearity via
collective enhancement and open the route to nonlinear optics with
single photons.

{\bf Acknowledgments: work supported by Austrian Science Fund projects
P17709+I119 N16}

\end{document}